\PassOptionsToPackage{unicode}{hyperref}
\PassOptionsToPackage{hyphens}{url}
\PassOptionsToPackage{dvipsnames,svgnames,x11names}{xcolor}
\documentclass[
  12pt]{article}

\usepackage{amsmath,amssymb}
\usepackage{iftex}
\ifPDFTeX
  \usepackage[T1]{fontenc}
  \usepackage[utf8]{inputenc}
  \usepackage{textcomp} 
\else 
  \usepackage{unicode-math}
  \defaultfontfeatures{Scale=MatchLowercase}
  \defaultfontfeatures[\rmfamily]{Ligatures=TeX,Scale=1}
\fi
\usepackage{lmodern}
\ifPDFTeX\else  
\fi
\IfFileExists{upquote.sty}{\usepackage{upquote}}{}
\IfFileExists{microtype.sty}{
  \usepackage[]{microtype}
  \UseMicrotypeSet[protrusion]{basicmath} 
}{}
\makeatletter
\@ifundefined{KOMAClassName}{
  \IfFileExists{parskip.sty}{%
    \usepackage{parskip}
  }{
    \setlength{\parindent}{0pt}
    \setlength{\parskip}{6pt plus 2pt minus 1pt}}
}{
  \KOMAoptions{parskip=half}}
\makeatother
\usepackage{xcolor}
\makeatletter
\ifx\paragraph\undefined\else
  \let\oldparagraph\paragraph
  \renewcommand{\paragraph}{
    \@ifstar
      \xxxParagraphStar
      \xxxParagraphNoStar
  }
  \newcommand{\xxxParagraphStar}[1]{\oldparagraph*{#1}\mbox{}}
  \newcommand{\xxxParagraphNoStar}[1]{\oldparagraph{#1}\mbox{}}
\fi
\ifx\subparagraph\undefined\else
  \let\oldsubparagraph\subparagraph
  \renewcommand{\subparagraph}{
    \@ifstar
      \xxxSubParagraphStar
      \xxxSubParagraphNoStar
  }
  \newcommand{\xxxSubParagraphStar}[1]{\oldsubparagraph*{#1}\mbox{}}
  \newcommand{\xxxSubParagraphNoStar}[1]{\oldsubparagraph{#1}\mbox{}}
\fi
\makeatother

\usepackage{longtable,booktabs,array}
\usepackage{calc} 
\usepackage{etoolbox}
\makeatletter
\patchcmd\longtable{\par}{\if@noskipsec\mbox{}\fi\par}{}{}
\makeatother
\IfFileExists{footnotehyper.sty}{\usepackage{footnotehyper}}{\usepackage{footnote}}
\makesavenoteenv{longtable}
\usepackage{graphicx}
\makeatletter
\def\maxwidth{\ifdim\Gin@nat@width>\linewidth\linewidth\else\Gin@nat@width\fi}
\def\maxheight{\ifdim\Gin@nat@height>\textheight\textheight\else\Gin@nat@height\fi}
\makeatother
\setkeys{Gin}{width=\maxwidth,height=\maxheight,keepaspectratio}
\makeatletter
\def\fps@figure{htbp}
\makeatother

\makeatletter
\@ifpackageloaded{caption}{}{\usepackage{caption}}
\AtBeginDocument{%
\ifdefined\contentsname
  \renewcommand*\contentsname{Table of contents}
\else
  \newcommand\contentsname{Table of contents}
\fi
\ifdefined\listfigurename
  \renewcommand*\listfigurename{List of Figures}
\else
  \newcommand\listfigurename{List of Figures}
\fi
\ifdefined\listtablename
  \renewcommand*\listtablename{List of Tables}
\else
  \newcommand\listtablename{List of Tables}
\fi
\ifdefined\figurename
  \renewcommand*\figurename{Figure}
\else
  \newcommand\figurename{Figure}
\fi
\ifdefined\tablename
  \renewcommand*\tablename{Table}
\else
  \newcommand\tablename{Table}
\fi
}
\@ifpackageloaded{float}{}{\usepackage{float}}
\floatstyle{ruled}
\@ifundefined{c@chapter}{\newfloat{codelisting}{h}{lop}}{\newfloat{codelisting}{h}{lop}[chapter]}
\floatname{codelisting}{Listing}

\makeatother
\makeatletter
\@ifpackageloaded{caption}{}{\usepackage{caption}}
\@ifpackageloaded{subcaption}{}{\usepackage{subcaption}}
\makeatother

\ifLuaTeX
  \usepackage{selnolig}  
\fi
\usepackage[]{natbib}
\usepackage{bookmark}

\IfFileExists{xurl.sty}{\usepackage{xurl}}{} 
\hypersetup{
  pdftitle={Title},
  pdfauthor={Author 1; Author 2},
  pdfkeywords={3 to 6 keywords, that do not appear in the title},
  colorlinks=true,
  linkcolor={black},
  filecolor={Maroon},
  citecolor={black},
  urlcolor={black},
  pdfcreator={LaTeX via pandoc}}

\newcommand{\anon}{1}

\newcommand{\etal}{{ et al. }}

\newcommand{\bGamma}{\boldsymbol \Gamma}
\newcommand{\hbeta}{\hat{\boldsymbol \beta}_{\mbox{\tiny \rm  F}}}
\newcommand{\halpha}{\hat{\boldsymbol\alpha}_{\mbox{\tiny \rm  F}}}
\newcommand{\htheta}{\hat{\boldsymbol\theta}_{\mbox{\tiny \rm  F}}}
\newcommand{\hW}{\hat{ \mathbf{W}}_{\mbox{\tiny \rm  F}}}
\newcommand{\btheta}{\boldsymbol \theta}
\newcommand{\bSigma}{\boldsymbol \Sigma}
\newcommand{\bbeta}{\boldsymbol \beta}
\newcommand{\balpha}{\boldsymbol \alpha}
\newcommand{\tbeta}{\tilde{\boldsymbol \beta}}

\newcommand{\bUpsilon}{\boldsymbol \Upsilon}

\newcommand{\X}{\mathbf{X}}
\newcommand{\Z}{\mathbf{Z}}
\newcommand{\V}{\mathbf{V}}
\newcommand{\sumn}{\sum_{i=1}^n}

\newcommand{\onen}{\frac{1}{n}}

\newcommand{\app}{\pi_{i}^{\text{app}}}
\newcommand{\appm}{\pi_{\delta i}^{\text{app}}}
\newcommand{\appsi}{\pi_{\delta i}^{\text{app}*}}

\newcommand{\WO}{\tilde{\mathbf{W}}_0}
\newcommand{\thetaO}{\tilde{\boldsymbol{\theta}}_0}
\usepackage{mathrsfs,amsmath,epsfig,amssymb,color,graphicx,bm,algpseudocode}
\usepackage{natbib,subcaption,hyperref,lineno}
\usepackage{algorithm}    
\usepackage{amsmath}    
\usepackage{amsfonts}   
\usepackage{booktabs}   
\usepackage{array}  
\newtheorem{theorem}{Theorem}
\newtheorem{remark}{Remark}

\graphicspath{{figures/}}
\date{}
\begin{document}

\def\spacingset#1{\renewcommand{\baselinestretch}%
{#1}\small\normalsize} \spacingset{1}


\if1\anon
{
  \title{\bf Efficient Poisson Subsampling for the Partially Linear Additive Cox  Model}
\author{ \small
	Dongxiao Han\textsuperscript{} \\
	\small NITFID, LPMC and KLMDASR, School of Statistics and Data Science, Nankai University\\
\small	Liuquan Sun\\
\small	SKLMS, Academy of Mathematics and Systems Science, Chinese Academy of Sciences,\\
\small	and School of Mathematical Sciences, University of Chinese Academy of Sciences
	\\
\small	Chunjie Wang\\
\small	School of Mathematics and Statistics, Changchun University of Technology\\
\small	Dehui Wang\\
\small	School of Mathematics and Statistics, Liaoning University\\
\small	HaiYing Wang\\
\small	Department of Statistics, University of Connecticut\\
\small	Haixiang Zhang\textsuperscript{$*$}\\
\small	School of Mathematics and KL-AAGDM, Tianjin University} 
  \maketitle
  \renewcommand{\thefootnote}{}
  \footnotetext[1]{$\dagger$Authors are listed in alphabetical order.}
\footnotetext[2]{$*$Corresponding author:  haixiang.zhang@tju.edu.cn}
  } \fi

\if0\anon
{
  \bigskip
  \bigskip
  \bigskip
  \begin{center}
    {\LARGE\bf Efficient Poisson Subsampling for the Partially Linear Additive Cox  Model}
\end{center}
  \medskip
} \fi

\bigskip
\begin{abstract}
To address the computational and storage challenges often encountered in
large-scale survival data analysis, we propose an efficient Poisson subsampling
method for  the partially linear additive Cox model.  This model provides a
flexible yet interpretable framework by incorporating linear covariate effects, additive nonparametric  components for  nonlinear covariates, and a nonparametric baseline hazard function.  The proposed method adopts B-spline basis functions to approximate the nonparametric components and employs the decorrelated score technique to construct a Poisson subsampling-based estimation equation, based on which we establish the asymptotic normality of the resulting  estimator and derive the optimal subsampling probabilities according to the L-optimality criterion. Furthermore, we design a two-step adaptive algorithm for practical implementation.   The proposed approach  enables computationally efficient statistical inference for large-scale survival analysis without processing the full dataset. We validate the performance of the
proposed method through extensive simulation studies and a real-world
application to a lymphoma cancer dataset, demonstrating its efficiency and
accuracy in large-scale settings.
\end{abstract}

\noindent%
\vfill

\newpage
\spacingset{1.8} 

\section{Introduction}\label{sec-intro}

With the rapid growth of data volume in modern application scenarios, traditional statistical methods frequently face computationally infeasible challenges, particularly in resource-constrained environments. Subsampling has emerged as a powerful and widely adopted technique in statistical and computational research. By carefully selecting a representative subset from the full dataset, subsampling allows for parameter estimation with reduced resource consumption while preserving the statistical validity of the resulting estimates. This makes it a practical solution for modern large-scale data challenges.

 Extensive research efforts have been devoted to the development of subsampling strategies for various statistical models. Specifically,  \cite{Wang2018-JASA} and \cite{Wang-MA2020} introduced  L-optimality based methods for logistic and quantile regression, respectively;  \cite{softmax-2021} studied the softmax regression via Poisson subsampling;
\cite{Func_sub-2023} investigated A-optimality methods for the functional generalized linear model; Shao et al. (\citeyear{PLM-JMLR-2025}) studied  the  partially linear model  by constructing an unbiased Neyman-orthogonal score function; \cite{Zhang-Wang2021}, Zuo et al. (\citeyear{Zuo-2021-CS});  \cite{Poisson-JASA-2021} proposed optimal subsample selection approaches in distributed computing environments, and \cite{Yu18122024} developed an online strategy for the streaming multinomial logistic model.    In high-dimensional settings, \cite{RCV-2024} provided a cross-validation procedure for selecting optimal subsamples. 
Comprehensive reviews of this field can be found in \citeauthor{OSP-review-2021} (\citeyear{OSP-review-2021}) and Yu et al. (\citeyear{SP-review}) and the references therein.

 Survival data arise frequently in many fields, such as biomedicine \cite[]{NMI-2024}, engineering \cite[]{Engineering2021}, and social sciences (Lee et al., \citeyear{Lee03092025}). Although recent years have seen the development of subsampling methods for such data, these methods are limited to settings where covariates have parametric effects.  For instance,  Zuo\etal(\citeyear{zuo2021-sim}) investigated the additive hazards model \cite[]{LIN1994}, and Zhang\etal(\citeyear{JCGS-Cox}) and Shao\etal(\citeyear{SHAO-Dcox}) studied the Cox model \cite[]{Cox1972}, with all these studies employing L-optimality subsampling;
 In distributed settings, Zhang\etal(\citeyear{DsubCox}) further proposed a divide-and-conquer based subsampling approach for the Cox model. The partially linear additive Cox (PLA-Cox; \citeauthor{PL-Cox}, \citeyear{PL-Cox}) model, which includes the  Cox model as a special case, integrates both parametric and nonparametric covariate effects.  Given its wide applicability, there is an urgent need for subsampling methods to alleviate the computational burden of fitting this model to large-scale data.

 In this article, we propose an efficient Poisson  subsampling method tailored for the PLA-Cox model. Our approach approximates nonparametric components using B-spline basis functions, constructs Poisson subsampling-based estimation equations via the decorrelated score technique, and derives  the optimal subsampling probabilities based on the L-optimality criterion. For practical implementation, we design a two-stage adaptive algorithm. In the first stage, a pilot subsample is drawn via uniform Poisson subsampling to obtain pilot estimators. These estimators are then used to determine the optimal subsampling probabilities for selecting the final subsample and estimating the target parameters in the second stage.  The differences between our work and prior studies on subsampling in survival analysis are as follows:
First, as mentioned above, existing subsampling methods are designed for models with fully parametric covariate effects (e.g., the additive hazards model and the Cox model), whereas our method targets the PLA-Cox model, which  includes both parametric and nonparametric covariate effects.
Second, we develop an efficient Poisson subsampling scheme  for the PLA-Cox model, which implements sampling without replacement (thus avoiding duplicate observations), distinguishing it from existing with-replacement subsampling strategies. Third, to address the nonparametric components of covariate effects, we adopt B-spline basis functions for approximation and employ the decorrelated score technique to eliminate the confounding influence of nuisance parameters. This distinguishes our estimation strategy from existing subsampling methods.  Moreover, the introduction of this nonparametric component substantially complicates the theoretical development, as it necessitates validating the B-spline approximation and adjusting for the decorrelation procedure.

 The remainder of this paper is outlined as follows: In
Section~\ref{sec-2}, we  develop a Poisson subsampling  method for the PLA-Cox model, along
with the asymptotic normality of the corresponding estimator. 
In
Section~\ref{sec-4}, we focus on determining the optimal subsampling
probabilities, and propose a two-stage algorithm that enables the adaptive
implementation of the optimal subsampling method.   In Section~\ref{sec-5}, we conduct simulation studies to examine the finite-sample performance of the proposed method.  An application to a real dataset is provided in Section~\ref{sec-6}.  Concluding remarks are presented in Section \ref{se-8}. All proofs are deferred to the
Supplementary Material.

\section{Poisson Subsampling-Based Estimation}\label{sec-2}

 Let $T$ and $C$ be the failure time and censoring time, respectively. Denote $\bm{X} = (X_1, \ldots, X_p)^\prime \in \mathbb{R}^p$ as the covariate vector for the parametric components, and $\bm{Z} = (Z_1, \ldots, Z_q)^\prime \in \mathbb{R}^q$  as the covariates for the nonparametric components. Conditional on $\mathbf{X}$ and $\mathbf{Z}$, we assume that $T$ and $C$ are independent. The observed time is defined as $Y = \min(T, C)$, with event indicator $\Delta = I(T \le C)$, where $I(\cdot)$ denotes the indicator function. The full dataset consists of $n$ independent and identically distributed observations from the population, denoted as $\mathcal{D}_n = \left\{ \left( \mathbf{X}_i, \mathbf{Z}_i, \Delta_i, Y_i
  \right) : i = 1, \dots, n \right\}$. The PLA-Cox model assumes
\begin{align}\label{PLcox}
  \lambda(t \mid \bm{X}, \bm{Z}) = \lambda_0(t) \exp\{\bm{\beta}_0^\prime \bm{X} + g(\bm{Z})\},
\end{align}
where $\lambda(t \mid \bm{X}, \bm{Z})$ is the conditional hazard function given $\mathbf{X}$ and $\mathbf{Z}$,  $\lambda_0(t)$ is an unspecified baseline hazard function,
$\bm{\beta}_0$ is a $p$-dimensional regression
parameter vector, and the nonparametric components $g(\mathbf{Z}) = \sum_{j=1}^q g_j(Z_j)$ consists of smooth functions of $Z_j$, the $j$th component of $\mathbf{Z}$, satisfying ${E}\{g_j(Z_j)\} = 0$ for $j = 1, \ldots, q$.

 We approximate each $g_j(\cdot)$ by using B-spline
basis functions. Suppose that each $Z_j$ takes values in $[a,b]$, where $a$ and $b$ are finite numbers.
Define $\mathcal{S}_\varrho$ as the space of polynomial
splines of degree $\varrho \geq 1$.  Consider a knot
sequence with $J$ interior knots
\begin{equation*}\label{knotseq}
  a=t_{-\varrho} = \ldots = t_{-1} = t_0  < t_1 < \ldots < t_J = t_{J+1} = \ldots = t_{J+\varrho+1}=b,
\end{equation*}
where $J$ may increase with sample size $n$.  
  For each $j = 1,\ldots,q$, we approximate $g_j(Z_j)$ by a linear combination of B-spline basis functions $\{B_{s,j}(Z_j)\}_{s=-\varrho}^{J}$ of order $\varrho + 1$.
The nonparametric components in model (\ref{PLcox}) can 
be approximated as $\bm{\alpha}^\prime \bm{B}(\bm{Z})$,
where
$\mathbf{B}(\mathbf{Z}) = \{B_{s,j}(Z_j), s=-\varrho, \ldots, J;
j=1,\ldots,q\}^\prime$ is the B-spline basis vector, and
$\balpha= \{\alpha_{s,j}, s=-\varrho, \ldots, J; j=1,\ldots,q\}^\prime$ is the
corresponding coefficient vector.

 The
negative log pseudo-partial likelihood function for the full data is
\begin{align*}
  \ell(\btheta)
  &= -\onen\sumn  \int_0^\tau \left[\bbeta^\prime \X_i + \balpha^\prime {\mathbf{B}}(\mathbf{Z}_i) -  \log\left\{n^{-1}\sum_{j=1}^n I(Y_j \geq t)\exp\{\bbeta^\prime \X_j + \balpha^\prime {\mathbf{B}}(\mathbf{Z}_j)\}\right\}  \right] dN_i(t),
\end{align*} 
where $\btheta = (\bbeta^\prime,\balpha^\prime)^\prime$, $\tau$ is a
prespecified positive constant, and
$N_i(t) = I(\Delta_i =1, Y_i \leq t)$ is a counting process, $i=1,\ldots,n$. However, when the full dataset is large, minimizing $\ell(\btheta)$ brings an enormous computational burden, severely limiting the practical application of the PLA-Cox model in massive data scenarios. 
We employ the Possion subsampling method to address this challenge.  Specifically, the $i$th observation is assigned a sampling probability $\pi_i$ satisfying $0 < \pi_i \leq r^{-1}$ with $\sum_{i=1}^n \pi_i = 1$, where $r$  represents the expected subsample size. Let
$u_1,\ldots,u_n$ be independent uniform random variables on $(0,1)$ that are
independent of the observed data.  For each individual $i$, we include the tuple
$(\mathbf{X}_i, \mathbf{Z}_i, \Delta_i, Y_i)$ in the subsample if
$u_i \leq r\pi_i$. This yields the subsampled dataset
$\mathcal{D}_{r^*} = \{(\mathbf{X}_i^*, \mathbf{Z}_i^*, \Delta_i^*, Y_i^*,
\pi_i^*)\}_{i=1}^{r^*}$, where for each observation $i$, $\mathbf{X}_i^* $ are the linear-effect covariates, $\mathbf{Z}_i^* $ are the nonparametric confounders, $\Delta_i^*$ is the event indicator, $\mathbf{Y}_i^*$ is the response variable, and $\pi_i^*$ is the sampling probability.
The realized subsample size $r^*$ satisfies
$\mathbb{E}(r^*) = r$. Based on $\mathcal{D}_{r^*}$, a  negative weighted log pseudo-partial
likelihood is given as follows:
\begin{eqnarray*}\label{sub-L1}
  \ell^*(\btheta)
  &=& -\frac{1}{n}\sum_{i=1}^{r^*}\frac{1}{r\pi_{i}^*}  \int_0^\tau \Big[\bbeta^\prime \X^*_i + \balpha^\prime {\mathbf{B}}(\mathbf{Z}^*_i)\nonumber \\
  &&-  \log\Big\{n^{-1}\sum_{j=1}^{r^*} (r\pi_{j}^{*})^{-1}I(Y^*_j \geq t)\exp\{\bbeta^\prime \X^*_j + \balpha^\prime {\mathbf{B}}(\mathbf{Z}^*_j)\}\Big\}  \Big] dN_i^*(t),
\end{eqnarray*}
where $N_{i}^*(t) = I(\Delta_{i}^* =1, Y^*_{i} \leq t)$,  $i=1,\ldots,r^*$.



 In the PLA-Cox model, \(\bm{\beta}\) is the target parameter requiring rigorous inference, whereas \(\bm{\alpha}\) is nuisance parameter which merely capture nonparametric confounding effects. As a result, we adopt the decorrelated score technique to eliminate the influence of \(\bm{\alpha}\) on \(\bm{\beta}\). Let
 \begin{eqnarray*} 
 \nabla_{\bbeta}\ell^*(\bbeta, \balpha) = {\partial\ell^*(\bbeta, \balpha) }/{\partial \bbeta},~~  \nabla_{\balpha}\ell^*(\bbeta, \balpha) ={\partial\ell^*(\bbeta, \balpha) }/{\partial \balpha}, 
 \end{eqnarray*} 
\begin{eqnarray*} 
\nabla_{\bbeta\balpha}\ell^*(\bbeta, \balpha) = {\partial^2\ell^*(\bbeta, \balpha) }/{\partial \bbeta\partial \balpha^\prime},~~\nabla_{\balpha\balpha}\ell^*(\bbeta, \balpha) ={\partial^2\ell^*(\bbeta, \balpha) }/{\partial \balpha\partial \balpha^\prime}, 
\end{eqnarray*} 
and
 $$\htheta = (\hbeta^\prime, \halpha^\prime)^\prime = \arg\min\limits_{\btheta} \ell(\btheta).$$
 If  $\htheta $ is available, we propose the following decorrelated score function
\begin{eqnarray*}\label{EQ-10}
  U^*(\bbeta;\halpha, \hW) = \nabla_{\bbeta}{ \ell^*(\bbeta, {\halpha}^{})} - \hW \nabla_{\balpha}{\ell^*(\bbeta, {\halpha}^{})},
\end{eqnarray*}
where  $\hW = \nabla^2_{\bbeta  \balpha}\ell(\htheta)\{{\nabla_{\balpha\balpha}^2 \ell(\htheta)}\}^{-1}$ is the projection matrix of $\nabla_{\bbeta}{ \ell^*(\hbeta, {\halpha}^{})}$ onto 
$\nabla_{\balpha}{\ell^*(\hbeta, {\halpha}^{})}$. 
Nevertheless, since $\halpha$ and $\hW$ are unavailable, we replace them with pilot estimators obtained from a pilot subsample.  Given the expected pilot subsample size $r_0$, we draw a pilot subsample \(\mathcal{D}_{r_0^*} = \left\{(\mathbf{X}_{i}^{0*}, \mathbf{Z}_{i}^{0*},  \Delta_{i}^{0*}, Y_{i}^{0*}, \pi_{i}^{0*})\right\}_{i=1}^{r_0^{*}}\) via uniform  Poisson subsampling with $\pi_i^{0*}=1/n$, $i=1,\ldots,r_0^{*}$. Based on this subsample, we compute \((\tilde{\boldsymbol{\beta}_0}, \tilde{\boldsymbol{\alpha}}_0) = \arg\min \ell^{0*}(\boldsymbol{\beta}, \boldsymbol{\alpha})\) and \(\tilde{\mathbf{W}}_0 = \nabla^2_{\boldsymbol{\beta}\boldsymbol{\alpha}} \ell^{0*}(\tilde{\boldsymbol{\beta}}_0, \tilde{\boldsymbol{\alpha}}_0) \big\{\nabla_{\boldsymbol{\alpha}\boldsymbol{\alpha}}^2 \ell^{0*}(\tilde{\boldsymbol{\beta}}_0, \tilde{\boldsymbol{\alpha}}_0)\big\}^{-1}\),  where
\begin{eqnarray*}
 \ell^{0*}(\bbeta, \balpha)
  &=& -\sum_{i=1}^{r_0^*}\frac{1}{r_0}  \int_0^\tau \Big[\bbeta^\prime \X^{0*}_i + \balpha^\prime {\mathbf{B}}(\mathbf{Z}^{0*}_i)\nonumber \\
  &&-  \log\Big\{\sum_{j=1}^{r_0^{*}} (r_0)^{-1}I(Y^{0*}_j \geq t)\exp\{\bbeta^\prime \X^{0*}_j + \balpha^\prime {\mathbf{B}}(\mathbf{Z}^{0*}_j)\}\Big\}  \Big] dN_i^{0*}(t),
\end{eqnarray*}
and $N_{i}^{0*}(t) = I(\Delta_{i}^{0*} =1, Y^{0*}_{i} \leq t)$, $i=1,\ldots,n$. Finally, we obtain the subsampling estimator \(\tilde{\boldsymbol{\beta}}\) by solving \(U^*(\boldsymbol{\beta}; \tilde{\boldsymbol{\alpha}}_0, \tilde{\mathbf{W}}_0) = 0\). 

 Our method inherits two core statistical advantages. First, from the computational perspective,  the proposed method significantly reduces the computational burden by only utilizing a small fraction of the full dataset, avoiding the high computational cost of full-data optimization. Second,  in terms of statistical inference, the orthogonality between the decorrelated score function and the gradient vector of the nuisance parameter \(\bm{\alpha}\)  eliminates the interference of \(\bm{\alpha}\) on \(\tilde{\boldsymbol{\beta}}\), ensuring valid inference for the target parameter $\bbeta_0$
 in the presence of nonparametric covariate effects.  Our approach differs from existing subsampling methods in survival analysis, which are designed exclusively for models with fully parametric covariate effects and thus cannot be directly extended to our setting involving both parametric and nonparametric covariate components.

 For any vector $\mathbf{u}$,  define $\mathbf{u}^{\otimes 0} = 1$,
$\mathbf{u}^{\otimes 1} = \mathbf{u}$ and
$\mathbf{u}^{\otimes 2} = \mathbf{u}\mathbf{u}^\prime$.  Let  
\begin{align*}
 S^{(k)} (t,\btheta) &= \onen \sumn I(Y_i \geq t) \V_i^{\otimes k}\exp(\btheta^\prime \V_i), \nonumber\\
 S_X^{(k)}(t,\btheta) &= \onen \sum_{i=1}^n I(Y_i\geq t) \X^{\otimes k}_i \exp(\btheta^\prime \V_i), \nonumber\\
 S_Z^{(k)}(t,\btheta) &= \onen\sum_{i=1}^n I(Y_i\geq t) \mathbf{B}^{\otimes k}(\Z_i) \exp(\btheta^\prime \V_i),\nonumber\\
 S_W^{(k)}(t,\hW,\btheta) &= \onen\sum_{i=1}^n I(Y_i\geq t)\{\X_i -
 \hW \mathbf{B}(\Z_i)\}(\X^{\otimes k}_i)^\prime \exp(\btheta^\prime
  \V_i),\nonumber\\
\intertext{and}
\mathbf{\Psi} &= \onen\sumn\int_0^\tau\left[\frac{S_W^{(1)}(t,\hW,\htheta)}{S^{(0)}(t, \htheta)} -  \frac{S_X^{(1)}(t,\htheta)S_W^{(0)}(t,\hW,\htheta)^\prime}{\{S^{(0)}(t, \htheta)\}^2} \right]dN_i(t),\nonumber
\end{align*}
 where $\mathbf{V}_i = (\mathbf{X}_i^\prime, \mathbf{B}(\mathbf{Z}_i)^\prime)^\prime$, $i=1,\ldots,n$ and $k$ = 0, 1.  The following conditions are required for establishing the asymptotic properties of $\tbeta$.
\begin{itemize}
\item[(C1)] $\int_0^\tau \lambda_0(t)dt < \infty$, and $P(T \geq \tau)>0$.
\item[(C2)] The covariate vector $\mathbf{X}$  is  bounded.
\item[(C3)] The limit of $\mathbf{\Psi}$ is nonsingular.
\item[(C4)] $\max_{1\leq i \leq n} (n\pi_i)^{-1} = O_P(1)$,   $r_0=o(n)$ and $r=o(n)$.
\end{itemize}

Condition (C1) is a regularity assumption in survival analysis. Condition (C2) is a boundedness constraint on $\mathbf{X}$, which is a
conventional requirement in the literature \cite[]{Huang-J2013-AOS}. This assumption maintains practical relevance for biomedical survival data, where typical covariates, such as treatment indicators, blood pressure measurements, age, and gender, are naturally bounded. Condition (C3) is imposed to guarantee the invertibility of the matrix $\mathbf{\Psi}$. Condition (C4) specifies that the minimal subsampling probability scales at order 1/n,  which is consistent with Assumption 5 in Wang et al. (\citeyear{IEEE-Poisson}). Moreover,  $r_0=o(n)$ and $r=o(n)$ ensure that the subsample sizes are much smaller than the full data size, a natural requirement to alleviate the computational burden. Theorem \ref{Th1} below  develops the asymptotic normality of  $\tbeta$.



\begin{theorem}\label{Th1}
 Under conditions~(C1)-(C4), conditional on
  $\mathcal{D}_n$ in probability, we have
  \begin{align*}
    \bSigma^{-1/2}(\tbeta - \hbeta) \stackrel{d}{\longrightarrow} \mathcal{N}(\mathbf{0},\mathbf{I}),
  \end{align*}
  where $\stackrel{d}{\longrightarrow}$ denotes convergence in distribution,  and
  $\bSigma =\mathbf{\Psi}^{-1}\mathbf{\Gamma}\mathbf{\Psi}^{-1}$ with
  \begin{align*}
    \mathbf{\Gamma} &=\frac{1}{n^2} \sumn \frac{1}{r\pi_i} \left[\int_0^\tau\left\{\X_i - \frac{S_X^{(1)}(t,\htheta)}{S^{(0)}(t,\htheta)} - \hW\left[\mathbf{B}(\Z_i) - \frac{S_Z^{(1)}(t,\htheta)}{S^{(0)}(t,\htheta)}\right]\right\}dM_i(t,\htheta)\right]^{\otimes 2},
  \end{align*}
and  $M_i(t,\btheta) = N_{i}(t) - \int_0^tI(Y_{i} \geq u)\exp(\btheta^\prime \V_{i})\lambda_0(u) du$, $i=1,\ldots,n$. 
\end{theorem}

 Next, we  aim to find the optimal probabilities that minimize the covariance matrix in Theorem \ref{Th1}.
Because $\boldsymbol{\Psi}^{-1}\boldsymbol{\Gamma}_1\boldsymbol{\Psi}^{-1} \preceq \boldsymbol{\Psi}^{-1}\boldsymbol{\Gamma}_2\boldsymbol{\Psi}^{-1}$ is equivalent to $\boldsymbol{\Gamma}_1 \preceq \boldsymbol{\Gamma}_2$ in the Loewner ordering,  this objective reduces to minimizing $\boldsymbol{\Gamma}$. However, since two positive definite matrices are not necessarily comparable, we instead minimize $\operatorname{tr}(\mathbf{\Gamma})$, a criterion justified by the fact that $\mathbf{\Gamma}_1 \preceq \mathbf{\Gamma}_2$ implies $\operatorname{tr}(\mathbf{\Gamma}_1) \leq \operatorname{tr}(\mathbf{\Gamma}_2)$. This strategy is known as the L-optimality criterion, which has been widely adopted in the field of subsampling \cite[]{Wang-MA2020}. We denote these optimal probabilities as $\{\pi_i^{\text{Lopt}}\}_{i=1}^n$, and their closed-form expressions are presented in Theorem~\ref{Th2}.

\begin{theorem}\label{Th2}
 Suppose that conditions~(C1)-(C4) are satisfied.
If the subsampling probabilities are chosen as
  \begin{align}\label{002}
    \pi_{i}^{\rm Lopt} =  \frac{\|\bGamma_i(\hW,\htheta)\|\wedge H}{\sum_{j=1}^{n} \{\|\bGamma_j(\hW,\htheta)\|\wedge H\}},~  \ \textrm{$i=1,\ldots,n$},
  \end{align}
  then $tr(\mathbf\Gamma)$ attains its minimum, where $a\wedge b = \min(a,b)$,
  \begin{eqnarray*}
  \bGamma_i(\hW,\htheta) & = & \|\int_0^\tau\left\{\X_i - \frac{S_X^{(1)}(t,\htheta)}{S^{(0)}(t,\htheta)} - \hW\left[\mathbf{B}(\Z_i) - \frac{S_Z^{(1)}(t,\htheta)}{S^{(0)}(t,\htheta)}\right]\right\}dM_i(t,\htheta)\|,\\
  H &=& \frac{\sum_{i=1}^{n - s}\bGamma_{(i)}(\hW,\htheta)_{}}{r - s},
  \end{eqnarray*}
  $\bGamma_{(1)}(\hW,\htheta) \leq \ldots \leq \bGamma_{(n)}(\hW,\htheta)$ are the  order statistics of $\bGamma_1(\hW,\htheta)$, $\ldots$, $\bGamma_n(\hW,\htheta)$,  and $s$ is an integer such that
  \begin{align*}
  \frac{\bGamma_{(n-s)}(\hW,\htheta)}{\sum_{i=1}^{n-s}\bGamma_{(i)}(\hW,\htheta)_{}} < \frac{1}{r - s},
  \end{align*}
and
  \begin{align*}
  \frac{\bGamma_{(n-s+1)}(\hW,\htheta)}{\sum_{i=1}^{n-s+1}\bGamma_{(i)}(\hW,\htheta)_{}} < \frac{1}{r - s+1},
  \end{align*}
with $\bGamma_{(n+1)}(\hW,\htheta) = \infty$.
\end{theorem}

\begin{remark}\label{Remark1}
 As shown by Wang\etal(\citeyear{IEEE-Poisson}), we can practically use $H = \infty$, and the performance of the estimator with $H = \infty$ is comparable to that obtained with the optimal sampling probabilities.
 This strategy has also been adopted in other Poisson subsampling-based methods (Wang\etal\citeyear{IEEE-Poisson}; \citeauthor{Poisson-JASA-2021} \citeyear{Poisson-JASA-2021}; Yao\etal\citeyear{softmax-2021}) and can improve computational efficiency.
\end{remark}

%
%

\section{Two-Step Subsampling  Algorithm}\label{sec-4}

 As shown in Section \ref{sec-2}, the L-optimal subsampling probabilities depend on several unobservable quantities: the full data estimator  \(\htheta \), the projection matrix \(\hW\), $\Lambda_0(t) = \int_0^t \lambda_0(s)ds$, \(S_X^{(1)}(t,\htheta)\), \(S_Z^{(1)}(t,\htheta)\), and \(S^{(0)}(t,\htheta)\). To overcome the unavailability of these quantities in advance, we propose the following two-step subsampling algorithm.


\textbf{Step 1: Pilot Subsampling and Hybrid Probability Calculation}. A pilot subsample $\mathcal{D}_{r_0^*}$ of expected size \(r_0 \ll n\) is drawn from \(\mathcal{D}_n\) via uniform Poisson subsampling. Based on this pilot subsample, we compute pilot estimators \(\thetaO = (\tilde{\boldsymbol{\beta}_0}, \tilde{\boldsymbol{\alpha}}_0)\) and \(\WO\).  We  estimate the cumulative baseline hazard 
$\Lambda_0(t)$ by the following Breslow estimator computed on the pilot subsample,
$$
\hat{\Lambda}_0^{\text{UNIF}}(t,\thetaO) = \sum_{i=1}^{r_0^*} \int_0^t\frac{dN_{i}^{0*}(u) }{\sum_{j=1}^{r_0^*} I(Y_j^{0*} \geq u) \exp(\thetaO^\prime \mathbf{V}_j^{0*})},
\label{pi_Brew}
$$
where $\mathbf{V}_i^{0*} = (\mathbf{X}_i^{0*\prime}, \mathbf{B}(\mathbf{Z}_i^{0*})^\prime)^\prime$,  $i=1,\ldots,n$.  In addition, we replace  \(S_X^{(1)}(t,\htheta)\), \(S_Z^{(1)}(t,\htheta)\), and \(S^{(0)}(t,\htheta)\) with their pilot subsample counterparts:
\begin{align*}
S_X^{0*(1)}(t,\thetaO) &= \frac{1}{r_0^{*}} \sum_{i=1}^{r_0^*} I(Y_i^{0*} \geq t) \mathbf{X}_i^{0*} \exp(\thetaO^\prime \mathbf{V}_i^{0*}), \\
S_Z^{0*(1)}(t,\thetaO) &= \frac{1}{r_0^{*}} \sum_{i=1}^{r_0^*} I(Y_i^{0*} \geq t) \mathbf{B}(\mathbf{Z}_i^{0*}) \exp(\thetaO^\prime \mathbf{V}_i^{0*}), \\
\intertext{and}
S^{0*(0)}(t,\thetaO) &= \frac{1}{r_0^{*}} \sum_{i=1}^{r_0^*} I(Y_i^{0*} \geq t) \exp(\thetaO^\prime \mathbf{V}_i^{0*}),
\end{align*}
which are computed  by using the pilot subsample.  Apply these substitutions to (\ref{002}), yielding the implementable optimal subsampling probabilities,
\begin{align*}
 \pi_{i}^{\text{app}}= \frac{\|\bGamma_i(\WO,\thetaO)\|}{\sum_{j=1}^{n} \|\bGamma_j(\WO,\thetaO)\|},~  \ \textrm{$i=1,\ldots,n$},
\end{align*}
where 
  \begin{eqnarray*}
  \bGamma_i(\WO,\thetaO) & = & \int_0^\tau\left\{\X_i - \frac{S_X^{0*(1)}(t,\thetaO)}{S^{0*(0)}(t,\thetaO)} - \WO\left[\mathbf{B}(\Z_i^{0*}) - \frac{S_Z^{0*(1)}(t,\thetaO)}{S^{0*(0)}(t,\thetaO)}\right]\right\}d\hat{M}_i(t,\thetaO),
  \end{eqnarray*}
with  $\hat{M}_{i}(t,\thetaO) = N_{i}(t) - \int_0^tI(Y_{i} \geq u)\exp(\thetaO^\prime
\V_{i})d\hat{\Lambda}_{0}^{\mbox{\tiny\rm UNIF}}(u,\thetaO)$, $i=1,\ldots,n$.

To address numerical instability caused by near-zero \(\pi_{i}^{\text{app}}\), we introduce a hybrid subsampling scheme: \(\appm = (1-\delta)\app + {\delta}/{n}\), where \(\delta \in (0,1)\) (e.g., \(\delta = 0.1\)) controls the proportion of mixture.  This scheme balances optimality and stability, where the term \((1-\delta)\app\) retains the optimal subsampling feature (selecting informative observations) and \({\delta}/{n}\)  guarantees a non-zero subsampling probability for all observations.

\textbf{Step 2: Final Subsampling and Estimator Computation}. We generate independent uniform random variables \(u_i \sim U(0,1)\) for  each observation $i$ in \(\mathcal{D}_n\), and include observation \(i\) in the final subsample \(\mathcal{D}_{r^*}\) if \(u_i \leq r\appm\), where \(r\) is the expected final subsample size. 
 The two-step Poisson subsampling estimator \(\hat{\bbeta}\)   is then computed by solving the decorrelated score equation
 \begin{align*}\label{SC-45}
U_{\delta}^*(\bbeta;\tilde{\boldsymbol{\alpha}}_0, \WO) = \frac{\partial \ell_{\delta}^*(\bbeta, \tilde{\boldsymbol{\alpha}}_0)}{\partial \bbeta} - \WO \frac{\partial \ell_{\delta}^*(\bbeta, \tilde{\boldsymbol{\alpha}}_0)}{\partial \balpha} =0,
  \end{align*}
  where $\ell_{\delta}^*(\bbeta, \balpha)$ has the same expression as
  $\ell_{}^*(\bbeta, \balpha)$  except that
  $r\pi_{i}^*$ is replaced with $\{(r\appsi)\wedge 1\}$.  Our proposed two-step subsampling method is summarized in the following Algorithm  \ref{algo1}.
 
%

%

\begin{algorithm}[htp]
\caption{ Poisson Subsampling Estimation}\label{algo1}
\noindent\textbf{Input}: 
The full data $\mathcal{D}_n = \{(\mathbf{X}_i, \mathbf{Z}_i, \Delta_i, Y_i)\}_{i=1}^n$; 
 the expected pilot subsample size $r_0$;  the expected final subsample size  $r$;  the proportion of mixture $\delta$.


\vspace{0.5em}

\noindent\textbf{Step 1: Pilot Subsampling and Hybrid Probability Calculation}
\begin{enumerate}
    \item  Draw a uniform pilot subsample: 
          \(\mathcal{D}_{r_0^*} = \left\{(\mathbf{X}_{i}^{0*}, \mathbf{Z}_{i}^{0*},  \Delta_{i}^{0*}, Y_{i}^{0*}, \pi_{i}^{0*})\right\}_{i=1}^{r_0^{*}}\),  where $r_0^*$ is the actual pilot subsample size.
    \item  Compute \(\thetaO\), \(\WO\)  and $\app$.
    \item Calculate hybrid subsampling probabilities:
          \[
          \appm = (1-\delta)\app + \frac{\delta}{n}. \label{PSP1}
          \]
\end{enumerate}

\noindent\textbf{Step 2: Final Subsampling and Estimator Computation}
\begin{enumerate}
    \item For each $i=1,\dots,n$:
          \begin{itemize}
              \item Generate $u_i \sim U(0,1)$;
              \item If $u_i \leq r  \appm$,  we include $(\mathbf{X}_i, \mathbf{Z}_i, \Delta_i, Y_i)$ in the subsample.
          \end{itemize}
    \item By solving the estimating equation  $U_{\delta}^*(\bbeta;\tilde{\boldsymbol{\alpha}}_0, \WO) = 0$,  we obtain the estimator $\hat{\boldsymbol{\beta}}$.
\end{enumerate}
\end{algorithm}

 We establish the asymptotic normality of 
$\hat{\bbeta}$ in the  next Theorem \ref{pros3}.

\begin{theorem} \label{pros3}
Assume conditions ~(C1)-(C4) hold.  Conditional on $\mathcal{D}_{n}$ and $\thetaO$,
  the two-step Poisson subsampling estimator $\hat{\bbeta}$ satisfies
  \begin{align*}
    \bSigma_{\delta}^{-1/2}(\hat{\bbeta}- \hbeta)
    \stackrel{d}{\longrightarrow} \mathcal{N}(0,\mathbf{I}),
  \end{align*}
  where $\bSigma_{\delta} =\mathbf{\Psi}^{-1}\mathbf{\Gamma}_{\delta}\mathbf{\Psi}^{-1}$ and
    \begin{align*}
    \mathbf{\Gamma}_{\delta} &=\frac{1}{n^2} \sumn \frac{1}{ \{(r\appm)\wedge 1\}} \left[\int_0^\tau\left\{\X_i - \frac{S_X^{(1)}(t,\htheta)}{S^{(0)}(t,\htheta)} - \hW\left[\mathbf{B}(\Z_i) - \frac{S_Z^{(1)}(t,\htheta)}{S^{(0)}(t,\htheta)}\right]\right\}dM_i(t,\htheta)\right]^{\otimes 2}.
  \end{align*}
\end{theorem}

For the purposes of conducting statistical inference, we estimate the covariance matrix of $\hat{\bbeta}$ by
\begin{align*}
  \hat{\mathbf{\Omega}}  = \hat{\mathbf{\Psi}}^{-1}\hat{\mathbf{\Gamma}}\hat{\mathbf{\Psi}}^{-1},
\end{align*}
 where
  \begin{align*}
    \hat{\mathbf{\Psi}} = \frac{1}{n}\sum_{i=1}^{r^*} \frac{\Delta_{i}^*}{\{(r\appsi)\wedge 1\}}\left[\frac{S_W^{*(1)}(t,\WO,\hat{\btheta})}{S^{*(0)}(t, \hat{\btheta})} -  \frac{S_W^{*(0)}(t,\WO,\hat{\btheta})S_X^{*(1)}(t,\hat{\btheta})}{\{S^{*(0)}(t, \hat{\btheta})\}^2} \right],
  \end{align*}
  and
  \begin{align*}
    \hat{\mathbf{\Gamma}} &=\frac{1}{n^2} \sum_{i=1}^{r^*}  \frac{1}{\{(r\appsi)\wedge 1\}^2} \left[\int_0^\tau\left\{\X^*_i - \frac{S_X^{*(1)}(t,\hat{\btheta})}{S^{*(0)}(t,\hat{\btheta})} - \WO\left[\mathbf{B}(\Z^*_i) - \frac{S_Z^{*(1)}(t,\hat{\btheta})}{S^{*(0)}(t,\hat{\btheta})}\right]\right\}d\hat{M}^*_i(t,\hat{\btheta})\right]^{\otimes 2},
  \end{align*}
together with $\hat{\btheta} = (\hat{\bbeta}^{\prime}, \tilde{\boldsymbol{\alpha}}_0^\prime)^\prime$, $d\hat{M}^*_{i}(t,\btheta) = dN_{i}^*(t) - I(Y^*_{i} \geq
t)\exp(\btheta^\prime \V^*_{i})d\hat{\Lambda}_{0}^{\mbox{\tiny\rm
    UNIF}}(t,\btheta)$,
\begin{align*}
S_X^{*(1)}(t,\btheta) &= \frac{1}{n}\sum_{i=1}^{r^*} \frac{1}{\{(r\appsi)\wedge 1\}} I(Y^*_i\geq t) \X^*_i \exp(\btheta^\prime \V^*_i),\\
S_Z^{*(1)}(t,\btheta) &= \frac{1}{n} \sum_{i=1}^{r^*}\frac{1}{\{(r\appsi)\wedge 1\}} I(Y^*_i\geq t) \mathbf{B}(\Z^*_i) \exp(\btheta^\prime \V^*_i),\\
S^{*(0)} (t,\btheta) &= \frac{1}{n} \sum_{i=1}^{r^*} \frac{1}{\{(r\appsi)\wedge 1\}} I(Y^*_i \geq t) \exp(\btheta^\prime \V^*_i),\\
\intertext{and}
S_W^{*(k)}(t,\mathbf{W},\btheta) &= \frac{1}{n}\sum_{i=1}^{r^*} \frac{1}{\{(r\appsi)\wedge 1\}}I(Y^*_i\geq t)\{\X^*_i - \mathbf{W} \mathbf{B}(\Z^*_i)\}(\X^{*\otimes k}_i)^\prime \exp(\btheta^\prime \V^*_i),
\end{align*}
$k=0,1$.

 The proposed two-step subsampling  algorithm enables data-dependent sampling probabilities and adapts to the intrinsic features of complex large-scale survival datasets.
Notably,  the covariance estimator $\hat{\mathbf{\Omega}}$ is merely constructed from the final subsample, which effectively alleviates computational burden while providing reliable uncertainty quantification for $\hat{\bbeta}$ to support statistical inference.

\section{Numerical Simulations}\label{sec-5}

In this section, simulation studies  were conducted to evaluate the estimation accuracy and computational efficiency of the proposed Poisson subsampling estimator.   Let $\lambda_0(t) = 0.5t$, $\boldsymbol{\beta}_0 = (-1, -0.5, 0, 0.5, 1)^\prime$ and $n=10^6$. We considered two distinct data-generating processes for covariate $\mathbf{X}$:
\begin{itemize}
    \item[] \textit{Case} \uppercase\expandafter{\romannumeral 1}:  $\X_{}$ followed a multivariate normal distribution $\mathcal{N}(\mathbf{0},\bUpsilon)$, where $\Upsilon_{js}$  denotes the 
$(j,s)$ entry of $\bUpsilon$, with $\Upsilon_{js}=1$ if $j=s$ and $\Upsilon_{js}=0.3$ if $j\neq s$.
    \item[] \textit{Case} \uppercase\expandafter{\romannumeral 2}:  $\X_{}$ followed a mixed multivariate normal distribution $0.5\mathcal{N}(-\mathbf{1},\tilde{\bUpsilon})+0.5\mathcal{N}(\mathbf{1},\tilde{\bUpsilon})$, where $\tilde{\Upsilon}_{js}=0.5^{|j-s|}$.
\end{itemize}

 Let $\mathbf{Z} = (Z_1, Z_2)^\prime$, with $Z_1$ and $Z_2$  independently drawn from a uniform distribution on \((0,1)\). The nonparametric functions were specified as $g_1(z) = \sin(2\pi z)$ and $g_2(z) = 5z^4 + 3z^2 - 2$.  The interior knots were $\{0.2, 0.3, 0.5, 0.6, 0.8\}$.
  The censoring time $C$ was simulated from a uniform distribution on $(0, c_0)$, where $c_0$ was selected to give a certain censoring rate (CR).   We set CR at
 $20\%$ and $60\%$, respectively. All the simulation results presented below
 are based on 1000 replications.

\subsection{Target 1: Sensitivity to  the Pilot Subsample Size $r_0$}
We began by examining the effect of  the pilot subsample size $r_0$ under   $\delta = 0.1$ and $\varrho = 3$. Define $ESE_j$ as the empirical standard error of estimator $\hat{\beta}_j$ computed over 1000 simulation repetitions,  where $\hat{\beta}_j$ is the $j$th component of  $\hat{\bbeta}$, $j=1,\cdots,p$. The average standard error (ASE) was then given by $ASE = \sum_{j=1}^p ESE_j/p$. Table~\ref{tab:1} displays ASE for the subsampling estimator across pilot subsample sizes $r_0 \in \{500, 600, 700\}$  and final subsample sizes $r \in \{600, 800, 1000\}$. 
From Table~\ref{tab:1}, 
 the ASE  shows minimal fluctuations as $r_0$ increases from 500 to 700. For example, in Case I with CR=20\% and  the final subsample size $r=1000$,  the ASE is $0.0538$ when $r_0 = 500$, and decreases only slightly to $0.0535$ for $r_0 = 600$ and $r_0 = 700$, corresponding to a maximum reduction of just $0.56\%$.
This indicates that $r_0=500$ is sufficient to generate stable pilot estimators, as further increasing $r_0$ does not meaningfully improve the precision of $\hat{\boldsymbol{\beta}}$. In addition, as expected, the ASE consistently decreases with the increase of  the subsample size $r$, while  the higher censoring rate (CR=60\%) generally leads to a larger ASE  compared to the lower censoring rate (CR=20\%).

%
%
%

\begin{table}[H]
  \begin{center}
    \caption{The ASE of the subsampling estimator with different pilot subsample size $r_0$.}
    \label{tab:1}
    \vspace{0.1in} 
    \begin{tabular}{lccccccccccc}
      \hline
      & &  & \multicolumn{3}{c}{CR=$20\%$} &  & \multicolumn{3}{c}{CR=$60\%$} \\
      \cline{4-6}\cline{8-10}
      & $r_0$ & &$r=600$ &$r=800$ &$r=1000$ & &$r=600$ &$r=800$ &$r=1000$\\
      \hline
      Case I
& 500 &&  0.0733 & 0.0616 & 0.0538 & & 0.0883 & 0.0748 & 0.0658 \\
& 600 && 0.0720 & 0.0607 & 0.0535 & & 0.0865 & 0.0740 & 0.0651 \\
& 700 && 0.0714 & 0.0603 & 0.0535 & &  0.0858 & 0.0729 &  0.0646 \\
      Case II
& 500 && 0.0725 & 0.0619 &  0.0543 & & 0.1117 & 0.0815 &  0.0696 \\
& 600 && 0.0715 & 0.0611 &  0.0538 & & 0.0871 & 0.0730 &  0.0645 \\
& 700 && 0.0708 & 0.0603 &  0.0528 & & 0.0858 & 0.0721 &  0.0633 \\
      \hline
    \end{tabular}
  \end{center}
\end{table}

\subsection{Target 2: Sensitivity to the Mixing Parameter $\delta$}
We analyzed how ASE values vary with $\delta$ under  $r_0 = 500$ and $\varrho = 3$. The approximate sampling distribution $\boldsymbol{\pi}_{\delta}^{\mathrm{app}}$ exhibits two boundary behaviors: it converges to the optimal subsampling distribution as $\delta$ approaches 0, and approaches the uniform subsampling distribution when $\delta$ approaches 1. 
Table ~\ref{tab:2} shows that 
$\delta=0.1$ achieves the lowest ASE across all scenarios, striking the optimal balance between optimality and stability. For example, in Case I with a 20\% censoring rate and 
$r=1000$, its ASE of 0.0538 is about 1.5\% lower than that of 
$\delta=0$ and 1.3\% lower than that of 
$\delta=0.3$.  Thus,
$\delta=0.1$ eliminates the numerical instability caused by near zero probabilities without sacrificing efficiency. As 
$\delta$ increases to 0.3 or 0.5, the ASE generally rises, reflecting a loss of optimality as the sampling distribution becomes too uniform. The advantage of 
$\delta=0.1$ is greater under challenging conditions such as a high censoring rate of 60\% or asymmetric covariates as in Case II, where stability is critical. We therefore recommend 
$\delta=0.1$  for practical implementation of our method.

\begin{table}[htp]
  \begin{center}
    \caption{The ASE of the subsampling estimator with different mixing rate $\delta$.}
    \label{tab:2}
    \vspace{0.1in} 
    \begin{tabular}{lcccccccccccccccc}
      \hline
      & &  & \multicolumn{4}{c}{CR=$20\%$} &  & \multicolumn{4}{c}{CR=$60\%$} \\
      \cline{4-7}\cline{9-12}
& $r$ & &$\delta=0$ &$\delta=0.1$  &$\delta=0.3$ &$\delta=0.5$ & & $\delta=0$ &$\delta=0.1$ &$\delta=0.3$ &$\delta=0.5$\\
      \hline
      Case I
& 600 &&  0.0740 &  0.0713  &  0.0718 &  0.0723   &  & 0.0907 &  0.0883  & 0.0887 &   0.0939 \\
& 800 && 0.0629 &  0.0616  &   0.0609 & 0.0624   &  & 0.0768 &  0.0748  &  0.0758 &  0.0787 \\
&1000 &&  0.0546 &  0.0538  & 0.0545 &  0.0556   &  &0.0682 & 0.0658  &  0.0669 &  0.0698 \\
      Case II
& 600 &&  0.0729 &  0.0725  &  0.0722 &  0.0734   &  & 0.1139 &  0.1017  &   0.1002 &  0.1066 \\
& 800 &&  0.0620 &  0.0619  &  0.0609 &  0.0621   &  & 0.0996 &  0.0815  &  0.0811 &  0.0845 \\
&1000 &&  0.0550 &  0.0543  &  0.0547 &  0.0549   &  & 0.0861 &  0.0696  &   0.0727 &  0.0755 \\
      \hline
    \end{tabular}
  \end{center}
\end{table}

\subsection{Target 3: Sensitivity to the Spline Degree $\varrho$}

Following the preceding investigation, we assessed the sensitivity of the proposed estimator to the spline degree 
$\varrho$ under 
$\delta = 0.1$ and 
$r_0 = 500$. The spline degree controls the flexibility of the B-spline basis used to approximate the nonparametric component 
$g(Z)$. A lower degree, such as 
$\varrho=3$ for a cubic spline, is computationally efficient but may underfit complex nonparametric relationships. A higher degree, such as 
$\varrho=5$ for a quintic spline, can capture more intricate patterns but risks overfitting and higher computational cost. Table ~\ref{tab:3} shows that the ASE values for 
$\varrho$=3,4 and 5 are nearly identical, with maximum differences of less than 5\%, showing that the spline degree has minimal impact on the estimation accuracy of 
$\hat{\boldsymbol{\beta}}$. For example, in Case I with a 20\% censoring rate and 
r=1000, the ASE value for 
$\varrho$ =3 is 0.0538, which is only 3.4\% lower than the value of 0.0558 for 
$\varrho$ =5. Given the minimal differences in ASE across 
$\varrho$ values, the robustness of the estimator to the spline degree, and the computational efficiency of lower degrees, $\varrho$=3 is the optimal choice.


\begin{table}[htp]
  \begin{center}
    \caption{The ASE of the subsampling estimator with different spline degree $\varrho$.}
    \label{tab:3}
    \vspace{0.1in} 
    \begin{tabular}{lcccccccccccccccc}
      \hline
      & &  & \multicolumn{3}{c}{CR=$20\%$} &  & \multicolumn{3}{c}{CR=$60\%$} \\
      \cline{4-6}\cline{8-10}
& $r$ & &$\varrho=3$ &$\varrho=4$ &$\varrho=5$ & & $\varrho=3$  &$\varrho=4$ &$\varrho=5$\\
      \hline
      Case I
& 600 &&  0.0733  &  0.0741 & 0.0753 &  &   0.0883  &  0.0915 & 0.1252 \\
& 800 &&  0.0616  & 0.0625 & 0.0631 &  &    0.0748  & 0.0777 & 0.0952 \\
&1000 &&   0.0538  & 0.0556 & 0.0558 &  &   0.0658  & 0.0687 & 0.0833 \\
      Case II
& 600 &&  0.0725  & 0.0736 & 0.0738 &  &   0.0892  & 0.0913 & 0.0996 \\
& 800 &&  0.0619  & 0.0620 & 0.0627 &  &    0.0815  &  0.0786 & 0.0877 \\
&1000 &&  0.0543  & 0.0548 & 0.0553 &  &   0.0696  & 0.0735 & 0.0838 \\
      \hline
    \end{tabular}
  \end{center}
\end{table}

\subsection{Target 4: Performance Comparison of Subsampling Methods}
To assess the performance advantage of our L-optimality criterion-based subsampling approach (Lopt), we compared it with uniform subsampling (UNIF) and the method of \cite{JCGS-Cox} (denoted as ``Oracle'').  We set $r_0 = 500$, $\delta=0.1$  and $\varrho = 3$.
Under the UNIF scheme, each observation receives equal subsampling probability $\pi_i = 1/n$ for $i=1,\dots,n$. The Oracle method assumes prior knowledge of the nonparametric functions $g_1(z)$ and $g_2(z)$. Performance evaluation employs four metrics: estimated bias (BIAS) computed as the sample mean of the estimates minus the full data estimator, empirical  standard error (SE), mean estimated standard error (ESE), and empirical 95\% coverage probability (CP) based on normal approximation.

\begin{table}[htp]
\begin{center}
 \caption{ Simulation results on  $\hat{\bbeta}_1$ with Case I and CR=$20\%$.}
 \label{tab:4}
\begin{tabular}{lllccccccccccccc}
\hline
 &Subsample Size  &    Methods        && BIAS & SE & ESE & CP    \\    
\hline
 & $r=600$ & Lopt     &&  -0.0008  &  0.0755 & 0.0778  & 0.958   \\
          && UNIF     &&  -0.0250  & 0.0858 &  0.0960  &0.964 \\ 
          && Oracle   &&  -0.0010 & 0.0765 & 0.0737  & 0.952  \\
 & $r=800$ & Lopt     &&  -0.0045  & 0.0619 & 0.0662  & 0.962    \\
          && UNIF     &&  -0.0167  &  0.0719 &  0.0827  & 0.974   \\ 
          && Oracle   &&  0.0024  & 0.0653 & 0.0629  & 0.942  \\ 
 & $r=1000$& Lopt     &&  -0.0030  & 0.0549 & 0.0587  & 0.978    \\
          && UNIF     &&  -0.0177  & 0.0638 & 0.0740  &  0.966   \\ 
          && Oracle   &&  0.0001  & 0.0559 & 0.0558  & 0.940    \\
\hline
\end{tabular}\\[2mm]
\end{center}
\end{table}

\begin{table}[htp]
\begin{center}
 \caption{ Simulation results on  $\hat{\bbeta}_1$ with Case I and CR=$60\%$.}
 \label{tab:S4}
\begin{tabular}{lllccccccccccccc}
\hline
 &Subsample Size  &    Methods        && BIAS & SE & ESE & CP    \\    
 & $r=600$ & Lopt     &&  0.0054  & 0.0910 & 0.0938  & 0.958   \\
          && UNIF     &&  -0.0329  &  0.1323 & 0.1429  & 0.958  \\ 
          && Oracle   &&  0.0096  & 0.0911  & 0.0904  & 0.944  \\
 & $r=800$ & Lopt     &&  0.0025  & 0.0773 & 0.0800  & 0.946   \\
          && UNIF     &&  -0.0023  &0.1134 & 0.1225  & 0.960   \\ 
          && Oracle   &&  0.0094  & 0.0785 & 0.0770  & 0.954  \\ 
 & $r=1000$& Lopt     &&  0.0038  & 0.0696 & 0.0710  &0.950   \\
          && UNIF     && -0.0228  & 0.0997 & 0.1093  & 0.952   \\ 
          && Oracle   && 0.0075  & 0.0685 & 0.0684  &0.938   \\
\hline
\end{tabular}\\[2mm]
\end{center}
\end{table}

Tables \ref{tab:4}-\ref{tab:S77} show that all three methods produce near‑zero BIAS values, implying the unbiasedness of the subsampling estimator for ${\boldsymbol{\beta}}_1$. The results for the other components of 
${\boldsymbol{\beta}}$ are similar and omitted. For Lopt, the ESE  closely matches the corresponding SE, demonstrating that the proposed covariance estimator 
$\hat{\mathbf{\Omega}}$ is reliable. For instance, at 
$r=1000$, SE = 0.0549 and ESE = 0.0587, a relative difference of only 6.9\%. In contrast, UNIF shows a larger discrepancy, as seen at 
$r=1000$ where SE = 0.0638 and ESE = 0.0740, a relative difference of 16.0\%. This is because uniform sampling ignores the information  of observations, leading to less accurate standard error estimation. This advantage of Lopt grows with the subsample size $r$, as more  high-information observations are retained.


\begin{table}[htp]
\begin{center}
 \caption{ Simulation results on  $\hat{\bbeta}_1$ with Case II and CR=$20\%$.}
\begin{tabular}{lllccccccccccccc}
\hline
 &Subsample Size  &    Methods        && BIAS & SE & ESE & CP    \\    
\hline
 & $r=600$ & Lopt     && -0.0117  & 0.0727 & 0.0745  & 0.960   \\
          && UNIF     && -0.0257  &0.0865  & 0.0924  & 0.956  \\ 
          && Oracle   &&  -0.0034  &0.0740 & 0.0722  & 0.948  \\
 & $r=800$ & Lopt     && -0.0090  & 0.0631 & 0.0634  & 0.944   \\
          && UNIF     && -0.0188  &0.0761 & 0.0794  & 0.956   \\ 
          && Oracle   &&   -0.0001  & 0.0634 &0.0617  & 0.942  \\ 
 & $r=1000$& Lopt     &&  -0.0005  & 0.0565 &0.0563  & 0.946  \\
          && UNIF     &&   -0.0159  &0.0672 & 0.0709  &0.950  \\ 
          && Oracle   &&  -0.0008 & 0.0530 & 0.0548  & 0.954   \\
\hline
\end{tabular}\\[2mm]
\end{center}
\end{table}

\begin{table}[htp]
\begin{center}
 \caption{ Simulation results on  $\hat{\bbeta}_1$ with Case II and CR=$60\%$$^\dagger$.}
\label{tab:S77}
\begin{tabular}{lllccccccccccccc}
\hline
 &Subsample Size  &    Methods        && BIAS & SE & ESE & CP    \\    
\hline
 & $r=600$ & Lopt     &&  -0.0066  & 0.0932 & 0.0915  & 0.958   \\
          && UNIF     &&  -0.0337  & 0.1287 & 0.1384  & 0.962  \\ 
          && Oracle   &&  0.0088  & 0.0891 & 0.0891  & 0.958  \\
 & $r=800$ & Lopt     && -0.0026  & 0.0791 &  0.0766  & 0.950  \\
          && UNIF     && -0.0241  & 0.1103 & 0.1181  & 0.960  \\ 
          && Oracle   && 0.0108  & 0.0738 &  0.0760  & 0.958 \\ 
 & $r=1000$& Lopt     &&-0.0027  & 0.0693 & 0.0678  & 0.958  \\
          && UNIF     &&-0.0165  &  0.0998 & 0.1053  & 0.962   \\ 
          && Oracle   &&  0.0098  &0.0659 & 0.0674  & 0.960   \\
\hline
\end{tabular}\\[2mm]
\end{center}
\end{table}

\begin{figure}[H]
  \centering
  \begin{subfigure}{0.38\textwidth}
    \includegraphics[width=\textwidth]{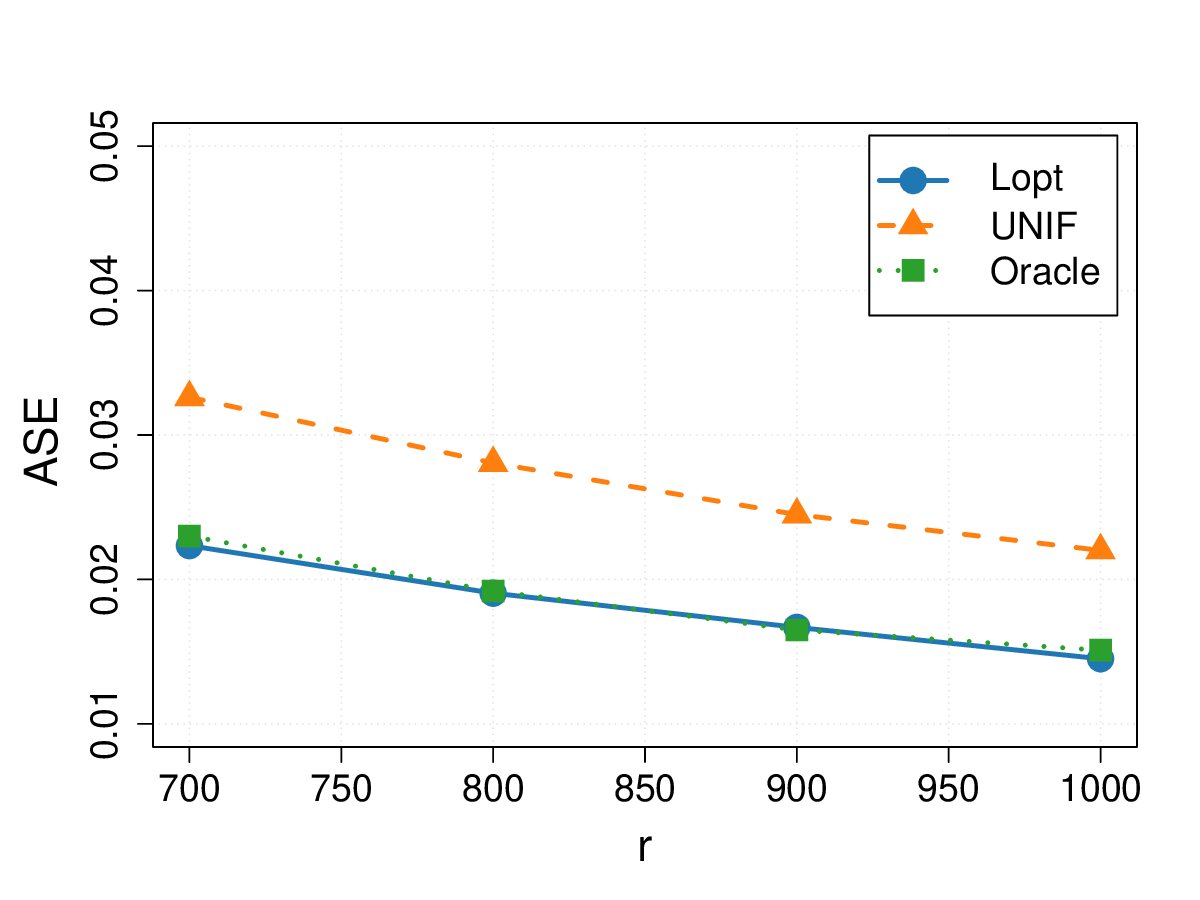}
    \caption{Case I with CR=$20\%$.}
  \end{subfigure}
  \begin{subfigure}{0.38\textwidth}
    \includegraphics[width=\textwidth]{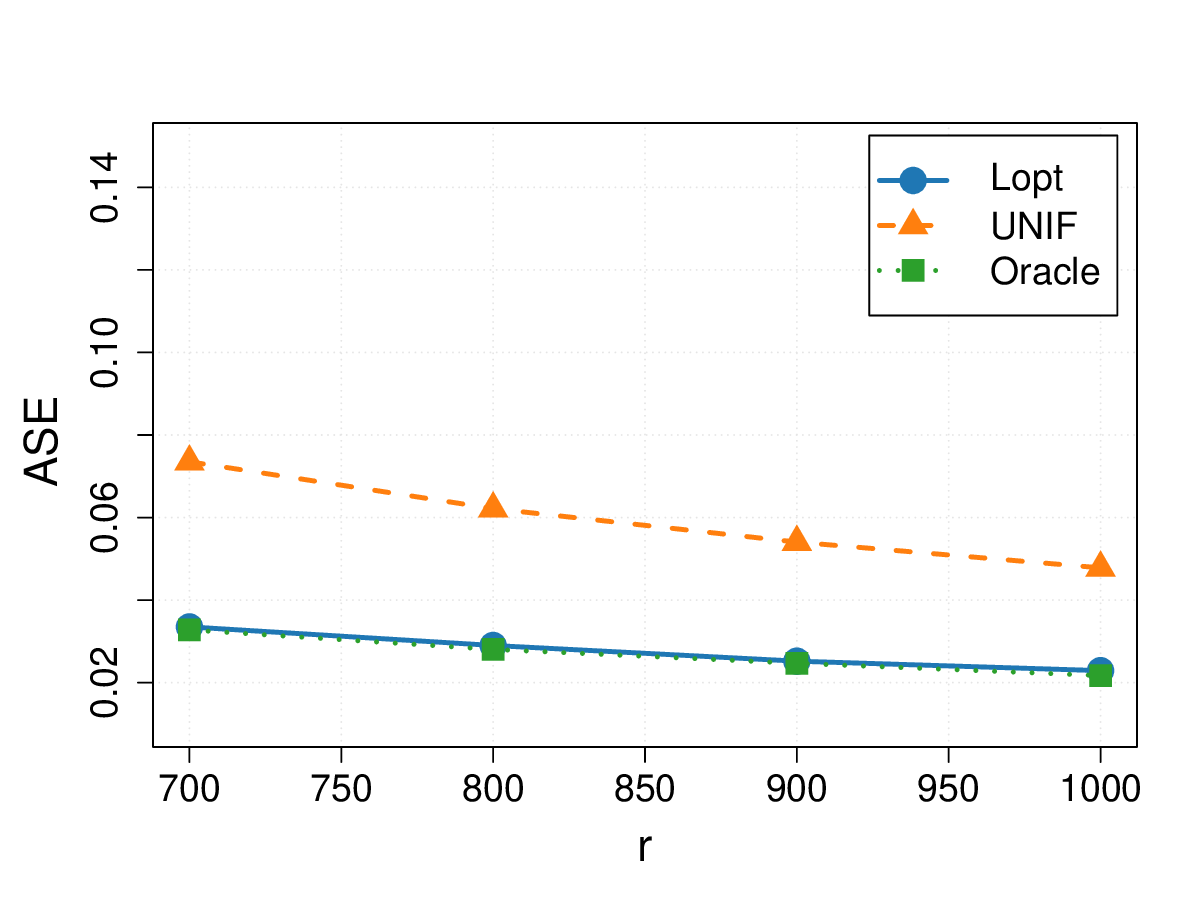}
    \caption{Case I with CR=$60\%$.}
  \end{subfigure}
    \begin{subfigure}{0.38\textwidth}
    \includegraphics[width=\textwidth]{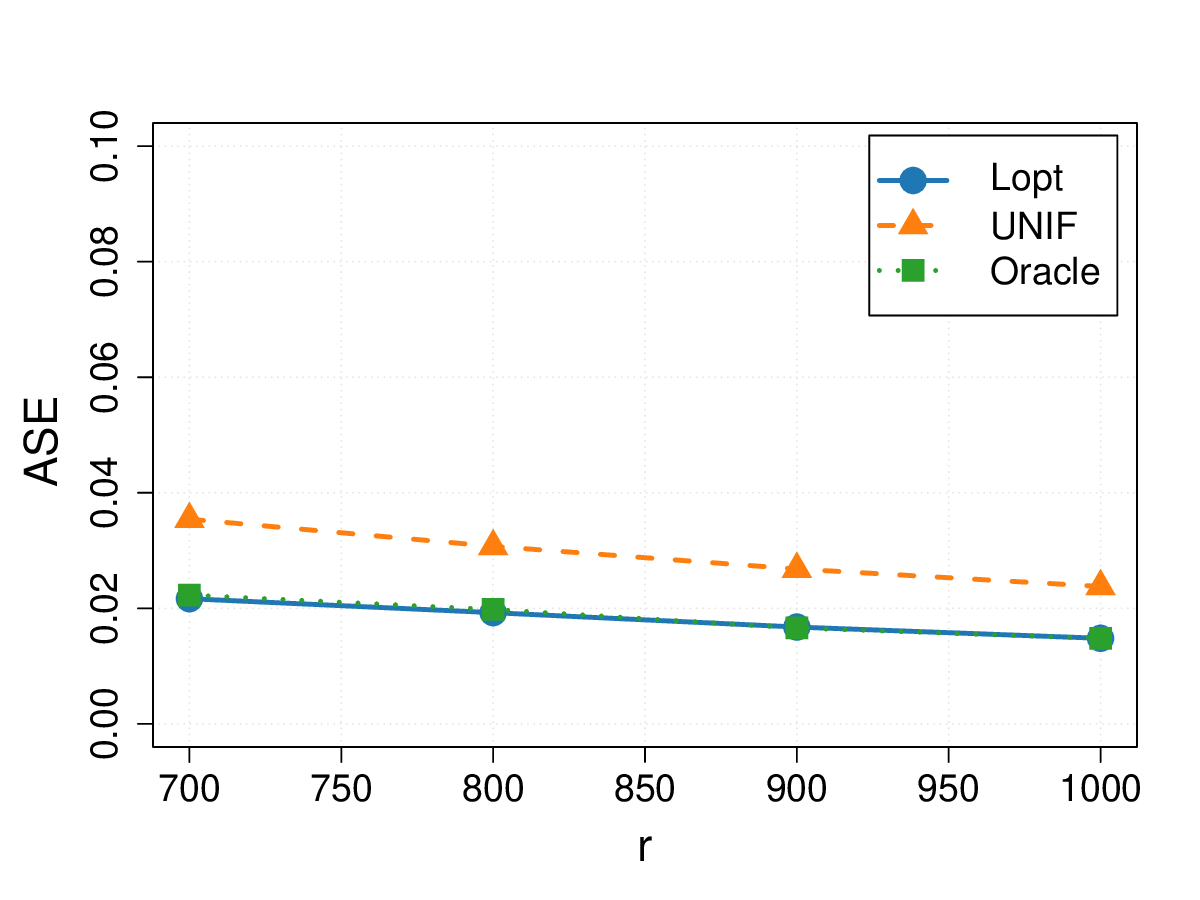}
    \caption{Case II with CR=$20\%$.}
  \end{subfigure}
  \begin{subfigure}{0.38\textwidth}
    \includegraphics[width=\textwidth]{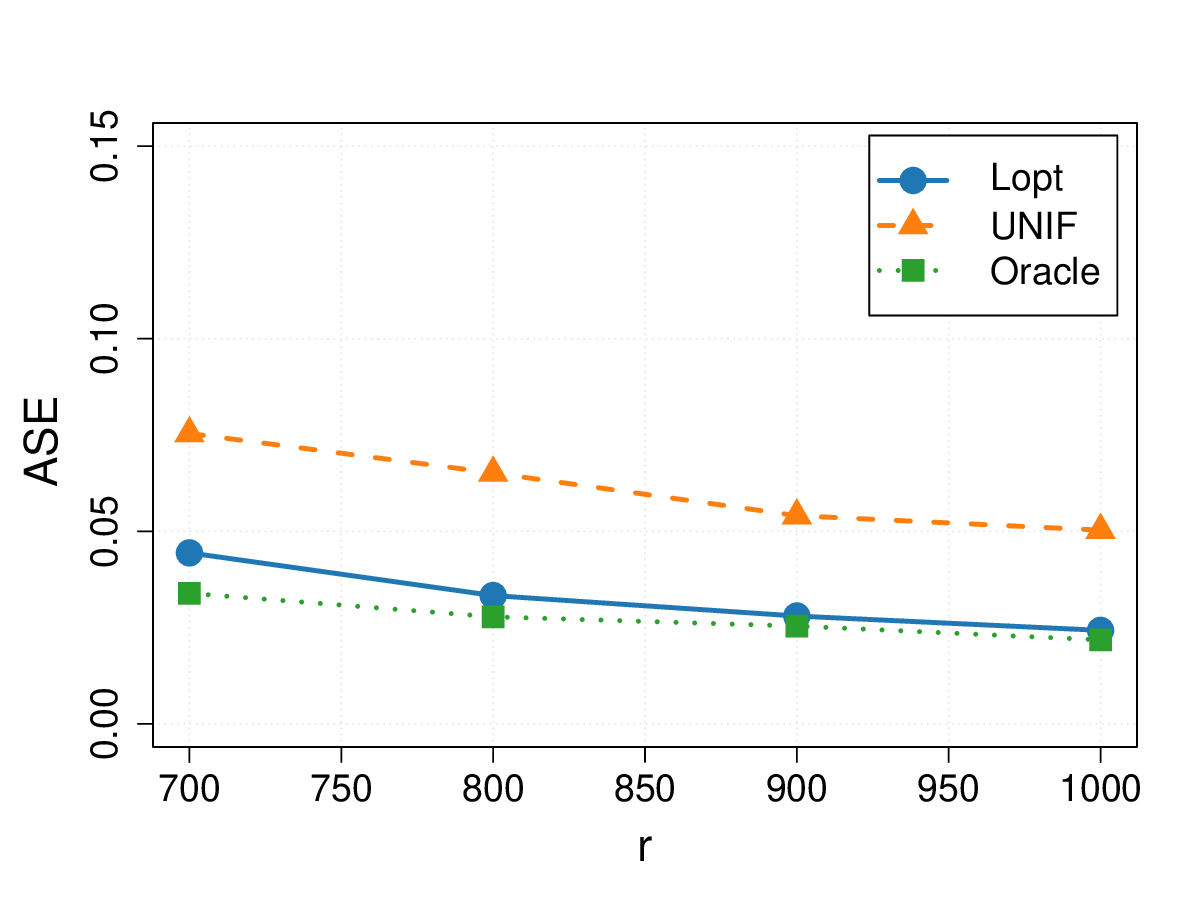}
    \caption{Case II with CR=$60\%$.}
  \end{subfigure}
 \vspace{-0.1cm}
\begin{center}
  \caption{ The ASEs for different subsampling methods.}
  \label{fig:1}
\end{center}
\end{figure}

Figure~\ref{fig:1} visualizes the ASE comparisons across all scenarios, further validating the superiority of Lopt. In all four subfigures, the ASE of Lopt is substantially lower than that of UNIF, and nearly identical to that of Oracle, especially when $r$ is large. For example, in Case I with CR=20\%, the ASE of Lopt at $r=1000$ is almost indistinguishable from Oracle, while UNIF’s ASE is significantly higher. This demonstrates that Lopt achieves near-optimal estimation accuracy without requiring prior knowledge of nonparametric functions.


\subsection{Target 5: Computational Efficiency Evaluation}
We conducted additional simulations to evaluate the computational efficiency of our proposed method. Data generation followed the Case I specification with covariates $\mathbf{X}$ constructed as previously described, except that the regression coefficient vector was \(\boldsymbol{\beta} = (0.5, \ldots, 0.5)'\) with dimensions $p = 5$ and $p = 50$.   We set $r_0 = 500$, $r=1000$, $\delta=0.1$  and $\varrho = 3$. All computations were performed in R on a laptop with 32GB RAM. We  report in Table~\ref{tab:7} the CPU times (in seconds) for computing $\hat{\boldsymbol{\beta}}$, averaged over 100 repetitions (excluding data generation time).
 From Table~\ref{tab:7} we see that 
both Lopt and UNIF achieve substantial computational savings compared to full data analysis. For example, when $p=5$ and $n=10^7$, the CPU time of Lopt is 7.459 seconds, which is only 4.2\% of that of full data analysis at 178.67 seconds.  This  implies that subsampling effectively reduces the computational burden by focusing on a small subset of informative observations, rather than processing the entire large dataset. UNIF is slightly faster than Lopt in all scenarios, which is attributed to the fact that UNIF does not require calculating optimal subsampling probabilities or allocation sizes.  While  UNIF offers faster computation, it suffers from a larger SE. In contrast,  Lopt balances computational efficiency and statistical performance.
The computational difference between subsampling methods and full data analysis widens with increasing 
$p$. This is because full data analysis requires processing all 
$n$ observations with 
$p$ covariates, leading to an increase in computational cost as 
$p$ grows, while subsampling methods only process a small subset of all observations, making the impact of $p$ on CPU time much smaller.



\begin{table}[H]
 \begin{center}
\caption{The CPU times of subsampling methods.}\label{tab:7}
\begin{tabular}{lllccccccccccccccc}
 \hline
      && Method && $n= 10^6$ && $n= 5\times 10^6$ && $10^7$ \\
      \hline
$p$=5&&Lopt    && 1.538   && 4.363   && 7.459  \\
      &&UNIF   && 0.982   && 1.259   && 1.494  \\
&& Full data   && 15.25   && 79.73   && 178.67  \\
$p$=50&&Lopt   && 12.906   && 28.844   && 44.644  \\
      &&UNIF   && 9.197   && 10.922   && 37.068 \\
&& Full data   && 58.95   && 292.14   && 628.01  \\
\hline
\end{tabular}\\
\end{center}
\end{table}

\section{A Real Data Example}\label{sec-6}

We illustrated the practical performance of the proposed two-step Poisson subsampling method by using a large-scale lymphoma cancer dataset from the Surveillance, Epidemiology, and End Results (SEER) program. The SEER is a authoritative and widely used source for population-level cancer survival research in the United States.
The dataset included complete records of \(N = 111,283\) lymphoma patients diagnosed between 1975 and 2007.
In this survival analysis, the follow-up time was censored at 60 months (5 years) after diagnosis.
Within the cohort, 46,067 individuals experienced the event of interest, leading to an empirical censoring rate of \(58.6\%\), which was consistent with the moderate-to-high censoring settings considered in our simulation studies.
The parametric covariate \(X\) was biological sex (coded as \(1\) for male and \(0\) for female), and the nonparametric nuisance covariate \(Z\) was age at diagnosis (centered and standardized).
The full data estimator using all observations was \(\hat{\beta} = 0.1956\), which served as the gold-standard benchmark for evaluating the subsampling estimator.

 We set \(r_0 = 500\), $\delta = 0.1$, \(\varrho=3\) and used interior knots at \(\{0.2, 0.3, 0.5, 0.6, 0.8\}\).
We generated \(1000\) independent subsamples at four incremental sizes: \(r = 600, 800, 1000, 1200\).
For each subsample size, we obtained both Lopt and UNIF estimators, and reported the point estimate, SE, ESE based on 1000 replications.
The results were summarized in Table~\ref{tab:real}.

\begin{table}[H]
  \begin{center}
    \caption{The results in the lymphoma cancer data.}\label{tab:real}
    \vspace{0.1cm}
    \small
    \begin{tabular}{lllccc}
      \hline
      & Method & Estimate   & SE       & ESE      \\
      \hline
      \(r=600\)
      & Lopt   & 0.18310    & 0.11297  & 0.10990  \\
      & UNIF   & 0.18983    & 0.12164  & 0.13277  \\
      \(r=800\)
      & Lopt   & 0.18353    & 0.09584  & 0.09434  \\
      & UNIF   & 0.18848    & 0.10632  & 0.11429  \\
      \(r=1000\)
      & Lopt   & 0.18548    & 0.08281  & 0.08395  \\
      & UNIF   & 0.18905    & 0.09645  & 0.10204  \\
      \(r=1200\)
      & Lopt   & 0.18684    & 0.07391  & 0.07648  \\
      & UNIF   & 0.19012    & 0.08991  & 0.09302  \\
      \hline
    \end{tabular}\\[2mm]
    \scriptsize
  \end{center}
\end{table}

Both Lopt and UNIF produce point estimates close to the full data estimator \(\hat{\beta}=0.1956\), confirming that subsampling-based inference is reliable for large-scale survival data.
Notably, the Lopt estimates are systematically closer to the full data estimator than UNIF across all subsample sizes.
As \(r\) increases from 600 to 1200, the Lopt estimates improve steadily from \(0.18310\) to \(0.18684\), gradually converging toward \(0.1956\).
In contrast, the UNIF estimates remain relatively unstable and deviate more noticeably from the benchmark.
The Lopt method achieves consistently smaller SE than UNIF at every subsample size.
For instance, at \(r=1000\), SE of Lopt is \(0.08281\), which is \(14.1\%\) lower than SE of UNIF (\(0.09645\)).
At \(r=1200\), the efficiency gain increases to \(17.8\%\).
This indicates that the L-optimal subsampling strategy effectively prioritizes informative observations, leading to more precise estimation than uniform sampling.

Moreover,  ESE closely matches the SE for Lopt across all subsample sizes, with relative differences below \(4\%\).
In contrast, ESE of UNIF noticeably overstates uncertainty and deviates from SE, especially at small subsample sizes.
This validates that the proposed covariance estimator is reliable  for practical applications.
Moreover, as the subsample size \(r\) increases, both methods improve in precision, with SE and ESE decreasing monotonically.
The Lopt method maintains its efficiency advantage across all levels of \(r\), demonstrating its stability and practical usefulness even with relatively small subsamples.


\section{Concluding Remarks}\label{se-8}

 In this article, we proposed an efficient Poisson subsampling method for the PLA-Cox model to address the computational and storage challenges of large-scale survival data analysis. The proposed method uses B-spline basis functions to approximate the additive nonparametric components and adopts the decorrelated score technique to construct a subsampling-based estimation equation. We established the asymptotic distribution of the resulting estimator and derived optimal subsampling probabilities under the L-optimality criterion. A two-step adaptive algorithm was developed for practical implementation. Extensive simulation studies and a real-world application to a lymphoma cancer dataset were conducted to validate the performance of the proposed method.

This research has several limitations that provide directions for future work. First, the hybrid subsampling scheme introduces a tuning parameter 
$\delta$,  whose selection is currently based on empirical experience (e.g., 
\(\delta = 0.1\)). Developing a data-driven method to select 
$\delta$ adaptively would further improve the method's robustness. Second, our framework focuses on independent and identically distributed datasets. Extending the Poisson subsampling framework to dependent survival data (e.g., clustered survival data or longitudinal survival data) is a valuable direction for future research. Third, the current method uses B-spline approximations for the nonparametric component 
\(g(\bm{Z})\). Exploring alternative nonparametric approximations (e.g., neural networks) may enhance the flexibility in capturing complex nonlinear relationships.

\vspace{0.5cm}
\noindent {\bf{AI Use Statement}}
During the preparation of this work, the authors use  Deepseek-V2 exclusively for English language editing and readability improvement.
\\
\noindent {\bf{Disclosure Statement}}
The authors report that there are no competing interests to declare.

\bigskip
\begin{center}
{\large\bf SUPPLEMENTARY MATERIAL}
\end{center}
\begin{description}
\item[] 
The supplementary material contains proofs of all the
theoretical results.
\end{description}

\section*{Acknowledgments}
Dongxiao Han's research was partly supported by the National Natural
Science Foundation of China (12471259 and 12231011), and Academy for Advanced Interdisciplinary Studies of Nankai University, and Tianjin Municipal Natural Science Foundation (23JCYBJC01270). Liuquan Sun's research was partially supported
by the National Natural Science Foundation of China (12571299).
Chunjie Wang's  research was partly supported by the National Natural
Science Foundation of China (12271060), and the Outstanding Youth Fund Project of Jilin Provincial Natural Science Foundation (20250101011JJ).
Dehui Wang's research was partly supported by the National Natural
Science Foundation of China (12271231, 1247012719).

\bibliography{bibliography.bib}

\end{document}